\documentclass[12pt,english]{article}
\usepackage{bm,bbm,lscape}
\usepackage{natbib, amsmath,authblk}
\usepackage{marvosym}
\usepackage{multicol}
\usepackage[british]{babel}
\usepackage{graphicx,xcolor,geometry,setspace,lineno}
\usepackage{pstricks}
\usepackage[british]{babel}
\usepackage[justification=justified,singlelinecheck=false,labelfont=it]{caption}
\usepackage{graphicx}
\usepackage{fancyhdr}
\definecolor{Mygrey}{gray}{0.65}

\title{Nonparametric inference in hidden Markov \\ models using P-splines}

  \author[1]{Roland Langrock}
  \author[2]{Thomas Kneib}
  \author[2]{Alexander Sohn}
  \author[1]{Stacy DeRuiter}

  \affil[1]{\small University of St Andrews, UK.}
  \affil[2]{\small Georg August University of G\"{o}ttingen, Germany}

 \date{}

\begin{document}

\begin{spacing}{1.2}
\maketitle

\vspace{-1em}

\begin{abstract}
\noindent
Hidden Markov models (HMMs) are flexible time series models in which the distributions of the observations depend on unobserved serially correlated states. The state-dependent distributions in HMMs are usually taken from some class of parametrically specified distributions. The choice of this class can be difficult, and an unfortunate choice can have serious consequences for example on state estimates, on forecasts and generally on the resulting model complexity and interpretation, in particular with respect to the number of states. We develop a novel approach for estimating the state-dependent distributions of an HMM in a nonparametric way, which is based on the idea of representing the corresponding densities as linear combinations of a large number of standardized B-spline basis functions, imposing a penalty term on non-smoothness in order to maintain a good balance between goodness-of-fit and smoothness. We illustrate the nonparametric modeling approach in a real data application concerned with vertical speeds of a diving beaked whale, demonstrating that compared to parametric counterparts it can lead to models that are more parsimonious in terms of the number of states yet fit the data equally well.
\end{abstract}

\vspace{0em}
\noindent
{\bf Keywords:} B-splines; cross-validation; forward algorithm; maximum likelihood; penalized smoothing.

\section{Introduction}\label{intro}

Due to their versatility and mathematical tractability, hidden Markov models (HMMs) have become immensely popular tools for modeling time series in which the observations depend on underlying nonobservable states. 
They have been applied in a diverse range of fields, and in particular in various biological scenarios, including DNA sequence analysis \citep{dur98}, scoring of sleep stages \citep{lan13sim}, 
mark-recapture studies \citep{pra05}, animal abundance estimation \citep{bor13}, animal behavior \citep{zuc08} and animal movement \citep{lan12e}. A basic $N$-state HMM involves two components: 1) an observed time series, which is typically referred to as the state-dependent process, since each of the corresponding observations is assumed to be generated by one of $N$ distributions as determined by the state of 2) an underlying (hidden) $N$-state Markov chain. A key property of HMMs is that dynamic programming algorithms can be used to efficiently evaluate the likelihood and to estimate the state sequence underlying the observations. For a comprehensive account of HMMs, see \citet{zuc09} and references therein.

In the literature, it is usually assumed that each state-dependent distribution is from a family of parametrically specified distributions. Choosing a family which is sufficiently flexible yet tractable can be difficult, for example if the {\em unknown} true state-dependent distributions are heavy-tailed, skewed or multi-modal. An unfortunate choice of the parametric family can lead, {\it inter alia}, to a poor fit, to biased estimates of the state transition probabilities, to poor predictive capacity and to wrong conclusions regarding the underlying system to be modeled. More specifically, parametric HMM formulations can lead to higher than adequate numbers of states being selected, for example by information criteria, simply because of the lack of flexibility of the considered state-dependent distributions in capturing the marginal distribution of the observations, potentially leading to state processes that are overly complex relative to the actual correlation structure. 

In a recent paper, \citet{yau11} suggested a nonparametric specification of the state-dependent distributions of an HMM for continuous-valued observations. Their technically challenging approach relies on Dirichlet process mixture priors that allow to specify a hyperprior on the space of potential probability distributions for the state-dependent distribution. \citet{dan12} developed an alternative frequentist approach based on the expectation-maximization algorithm, using log-concave densities or smoothing splines in the M-step in order to flexibly estimate the state-dependent distributions. He focused on the special case of two states, with one of the two state-dependent distributions modeled parametrically, arguing that this type of model is most relevant for applications and that computational and identifiability issues may arise in more difficult scenarios. However, it has recently been shown by \citet{gas13} and \citet{ale14} that identifiability in nonparametric HMMs holds under fairly weak conditions, which in practice will usually be satisfied, namely that the transition probability matrix of the unobserved Markov chain has full rank 
and that the state-dependent distributions are distinct. 

In this manuscript, we develop a novel nonparametric estimation approach involving an easy-to-implement and computationally feasible estimation algorithm. The main idea is to represent the densities of the state-dependent distributions as linear combinations of a large number of standardized B-spline basis functions, imposing a penalty term in order to arrive at an appropriate balance between goodness-of-fit and smoothness for the fitted densities. Each B-spline basis function is still associated with a separate parameter, leading to a model with finite-dimensional parameter space. However, the dimensionality is high and the separate parameters are not of interest. We therefore call our approach nonparametric, which is in line with the standard terminology in the statistical literature on smoothing methods, where (penalized) spline approaches are classified as nonparametric approaches \citep[see, for example,][]{rup03}. The nonparametric approach avoids assumptions regarding the form of the state-dependent distributions and hence is expected to be particularly useful, if only as an exploratory tool, in scenarios where the distribution of observations within states appears to be of a complicated form, making it hard to specify a suitable parametric family. 

We investigate practical issues involved when modeling the state-dependent distributions nonparametrically, both in simulations and by means of a real data case study, modeling vertical speeds of a diving Blainville's beaked whale ({\it Mesoplodon densirostris}). This species has been the focus of a considerable amount of research, motivated by mass strandings that were associated with naval sonar operations \citep{cox06}. They seem particularly sensitive to acoustic disturbance, altering their diving and foraging behavior in response to military sonar \citep{mcc11,tya11}. Quantitative description and comparison of normal and disturbed behavior are crucial to measuring the impact of anthropogenic noise, but are challenging given the diverse, sometimes subjective methods commonly used to classify whale dives and summarize dive behavior \citep{hoo01}. We illustrate that flexible nonparametric modeling of the state-dependent distributions of an HMM can be beneficial in this respect, and discuss some practical aspects particularly regarding potential advantages of the nonparametric approach compared to parametric HMM formulations.


\section{Nonparametric hidden Markov models}
\label{basics}


\subsection{Model formulation and penalized likelihood}
\label{hmms}

Let the observable state-dependent stochastic process be denoted by $\{ X_t \}_{t=1}^T$, and the underlying nonobservable $N$-state Markov chain by $\{ S_t \}_{t=1}^T$. We assume a basic dependence structure where given the current state of $S_t$, the variable $X_t$ is conditionally independent from previous and future observations and states, and where the Markov chain is of first order and homogeneous. Extensions to this basic structure are briefly discussed in Section \ref{discuss}. We summarize the probabilities of transitions between the different states in the $N \times N$ transition probability matrix (t.p.m.) $\boldsymbol{\Gamma}=\left( \gamma_{ij} \right)$, where $\gamma_{ij}=\Pr \bigl(S_{t}=j\vert S_{t-1}=i \bigr)$, $i,j=1,\ldots,N$.
The initial state probabilities are summarized in the row vector $\boldsymbol{\delta}$, where $\delta_{i} = \Pr (S_1=i)$, $i=1,\ldots,N$. 
For such an HMM, with observations $x_1,\ldots,x_T$ and parameter vector $\boldsymbol{\theta}$, the likelihood is given by
\begin{equation*}
\mathcal{L}^{\text{HMM}}(\boldsymbol{\theta}) = \sum_{s_1=1}^N \ldots \sum_{s_T=1}^N \delta_{s_1} \prod_{t=1}^T f (x_t | s_t)  \prod_{t=2}^T \gamma_{s_{t-1},s_t} \, .
\end{equation*}
In this form, the likelihood involves $N^T$ summands, rendering its evaluation infeasible even for a small number of states, $N$, and a moderate number of observations, $T$. However, the application of the recursive scheme called the {\em forward algorithm} leads to a much more efficient way of calculating the likelihood, via the matrix product expression
\begin{equation}\label{lik}
\mathcal{L}^{\text{HMM}}(\boldsymbol{\theta}) = \boldsymbol{\delta} \mathbf{Q}(x_1) \boldsymbol{\Gamma} \mathbf{Q}(x_{2}) \ldots \boldsymbol{\Gamma} \mathbf{Q}(x_{T}) \mathbf{1} \, ,
\end{equation}
where $\mathbf{Q}(x_t)= \text{diag} \bigl( f_1 (x_{t}), \ldots, f_N (x_{t}) \big)$, with $f_i(x_t) =f (x_{t} | S_{t}=i)$ denoting the density of the $i$-th state-dependent distribution, and where $\mathbf{1}\in \mathbbm{R}^N$ is a column vector of ones. The computational cost of evaluating (\ref{lik}) is {\it linear} in the number of observations, $T$, such that a numerical maximization of the likelihood becomes feasible in most cases. The popular alternative for parameter estimation in HMMs, namely usage of the expectation-maximization algorithm, is not considered in this work, since we agree with \citet{mac14} in there being no apparent reasons to prefer it over direct likelihood maximization. In our view, the direct maximization approach is more convenient to work with and more attractive to users, in particular since it has the crucial practical advantage that modifications in the model formulation usually require only very minor alterations in the code used to fit an HMM.

Here we are concerned with the nonparametric estimation of the densities $f_1, \ldots, f_N$, which we conduct following ideas from \citet{sch12}. More specifically, we suggest to represent each of these densities as a finite linear combination of basis functions $\phi_{-K},\ldots,\phi_{K}$, which are (known and fixed) probability density functions, as follows:
\begin{equation}\label{lincom2}
{f}_i(x) = \sum_{k=-K}^K {a}_{i,k} \phi_k(x) \, , \quad  i= 1,\ldots,N \, .
\end{equation}
Throughout this work, we use the same set of basis functions for each state-dependent distribution. Clearly, ${f}_i(x)$ is a probability density function if $\sum_{k=-K}^K {a}_{i,k} =1$ and ${a}_{i,k} \geq 0$ for all $k=-K,\ldots,K$. To enforce these constraints, the coefficients to be estimated, $a_{i,-K},\ldots,a_{i,K}$, are transformed using the multinomial logit link
$a_{i,k} = \exp(\beta_{i,k})/\{\sum_{j=-K}^K \exp(\beta_{i,j})\}$,
where we set $\beta_{i,0}=0$ for identifiability. In principle, any set of densities $\phi_{-K},\ldots,\phi_{K}$ can be used to approximate $f_i(x)$ as in (\ref{lincom2}). We follow \citet{sch12} and use (cubic) B-splines, in ascending order in the basis used in (\ref{lincom2}), with equally spaced knots and standardized such that they integrate to one. B-splines form a numerically stable, convenient basis for the space of polynomial splines, i.e., piecewise polynomials that are fused together smoothly at the interval boundaries; see \citet{deb78} and \citet{eil96} for more details. 
In most cases, cubic B-splines are a suitable default since they are twice continuously differentiable and therefore yield visually smooth density estimates.

The number of B-splines employed in the mixture specification (\ref{lincom2}) determines the potential flexibility. A larger number of basis elements will yield estimates that follow the data very closely but may be too wiggly, while a small number of basis elements yields very smooth estimates that may, however, be severely biased. To overcome the problem of selecting an optimal number of basis elements, we follow the penalized spline approach by \citet{eil96} and modify the log-likelihood by including a penalty on the sums of squared ($m$-th order) differences between coefficients associated with adjacent B-splines. Crucially, the characteristic HMM likelihood structure given in (\ref{lik}) remains valid, with the expression given in (\ref{lincom2}) applying to $f_1(x_t),\ldots,f_N(x_t)$ in the diagonal matrices $\mathbf{Q}(x_t)$, $t=1,\ldots,T$. For independent realizations $x_1,\ldots,x_T$, the corresponding penalized log-likelihood is given by
\begin{equation}\label{plik}
l_p^{\text{HMM}}(\boldsymbol{\theta},\boldsymbol{\lambda}) = \log \bigl( \mathcal{L}^{\text{HMM}}(\boldsymbol{\theta}) \bigr) - \left[ \sum_{i=1}^N \frac{\lambda_i}{2} \sum_{k=-K+m}^{K} \bigl( \Delta^m a_{i,k} \bigr)^2 \right] \, ,
\end{equation}
where $\Delta a_k = a_k-a_{k-1}$ and $\Delta^m a_k = \Delta (\Delta^{m-1} a_k)$ are difference operators, the parameter vector $\boldsymbol{\theta}$ comprises the state transition probabilities and the parameters $\beta_{i,k}$ ($i=1,\ldots,N$, $-K\leq k \leq K$, $k\neq 0$), and  $\boldsymbol{\lambda}=(\lambda_1,\ldots,\lambda_N)$ is a vector of smoothing parameters.
The penalty term penalizes roughness of the estimator, and the choice of $\boldsymbol{\lambda}$ determines how much emphasis is put on goodness-of-fit and on smoothness, respectively. In particular, choosing $\boldsymbol{\lambda}=(0,\ldots,0)$ leads to an unpenalized estimation, whereas for $\lambda_i\rightarrow\infty$, $i=1,\ldots,N$, the penalty will dominate the likelihood and for each $i$ we will obtain a sequence of weights $a_{i,k}$ that follow a polynomial of order $m-1$ in $k$. 
We will use $m=2$ in the remainder, since this provides an approximation to the integrated squared second derivative penalty that is popular in the context of smoothing splines.

The penalty term allows us to circumvent the problem of selecting an optimal number of basis elements, since it effectively reduces the number of free basis parameters and yields an adaptive fit to the data as long as the smoothing parameter is chosen in a data-driven way. Basically, we only have to ensure that the number of basis elements is large enough to provide enough flexibility for reflecting the structure of the state-dependent distributions. Once this threshold is passed, a further increase in the number of basis elements does no longer change the fit to the data much due to the impact of the penalty. Allowing for different smoothing parameters across states will be important in some circumstances, for example if the (true) densities for some state-dependent distributions are much more wiggly than for others, or if some states of the Markov chain are visited much less frequently than others, potentially requiring higher penalties on roughness due to less observations being available. 

\subsection{Model fitting and inference}


\subsubsection{Parameter estimation}

The penalized log-likelihood (\ref{plik}) can be maximized numerically, corresponding to a simultaneous estimation of the Markov chain parameters and the coefficients that determine the state-dependent distributions according to (\ref{lincom2}). Of the technical issues arising in the numerical maximization, discussed in detail in  \citet{zuc09}, 
the most important one is that of local maxima: particularly for complicated models, e.g., such with a relatively high number of states, it will sometimes happen that the numerical search fails to find the maximum penalized likelihood estimate, and returns a local maximum instead. The best way to address this issue seems to be to use a number of different sets of initial values, in order to maximize the chances of finding the global maximum.  

\subsubsection{Choice of the smoothing parameter vector}\label{cv}

Cross-validation techniques can be used for choosing the smoothing parameter. For a given time series, we suggest to generate $C$ random partitions such that in each partition a high percentage of the observations, e.g., 90\%, form the calibration sample, while the remaining observations constitute the validation sample. For each of the partitions and any given $\boldsymbol{\lambda}$, the model is then fitted (i.e., calibrated) using only the observations from the calibration sample (in a times series of exactly the same length as the original one, treating the data points from the validation sample as missing data; this is straightforward in the HMM framework, as detailed in \citealp{zuc09}). Subsequently, scoring rules can be used on the validation sample to assess the model for the given $\boldsymbol{\lambda}$ and the corresponding calibrated model. We consider the log-likelihood of the validation sample, under the model fitted in the calibration stage, as the score of interest (now treating the data points from the calibration sample as missing data in the time series).
From some pre-specified grid $\boldsymbol{\Lambda} \subset \mathbbm{R}_{\geq 0}^N$, we then select the $\boldsymbol{\lambda}$ that yields the highest mean score over the $C$ cross-validation samples. The number of samples $C$ needs to be high enough to give meaningful scores (i.e., such that the scores give a clear pattern rather than noise only), but must not be too high to allow for the approach to be computationally feasible. Furthermore, in scenarios where a full grid search over $\boldsymbol{\Lambda}$ is computationally infeasible, we suggest the following pragmatic algorithm for finding an appropriate $\boldsymbol{\lambda}$:
\begin{itemize}
\item[1.] choose an initial point $\boldsymbol{\lambda}_0^*$ from the grid $\boldsymbol{\Lambda}$ (and set $k=0$);
\item[2.] calculate the mean score for $\boldsymbol{\lambda}_k^*$ and each direct neighbor of $\boldsymbol{\lambda}_{k}^*$ on the grid;
\item[3.] from these values choose $\boldsymbol{\lambda}_{k+1}^*$ as the one that yielded the highest mean score;
\item[4.] repeat 2.\ and 3.\ until $\boldsymbol{\lambda}_{k+1}^*=\boldsymbol{\lambda}_k^*$.
\end{itemize}


\subsubsection{Uncertainty quantification}\label{se}

Uncertainty quantification, on both the estimates of the transition probabilities and on the estimates of the densities of the state-dependent distributions, can be performed using a parametric bootstrap. In particular, from the bootstrap replications one can obtain pointwise confidence intervals for the estimated densities as the corresponding quantiles at a specific point in the support. These pointwise confidence intervals can also be used to obtain simultaneous confidence bands for the complete density following \citet{kri10}. The idea is to rescale the pointwise confidence bands with a constant factor until a certain fraction of complete densities from the set of bootstrap replications is contained in the confidence band. By construction, these simultaneous bands use the pointwise intervals to assess local uncertainty about the estimated density and inflate this local uncertainty such that simultaneous coverage statements are possible.




\section{Simulation study}
\label{simuls}

\begin{figure}[!htb]
\begin{center}
{\includegraphics*[width=1\textwidth]{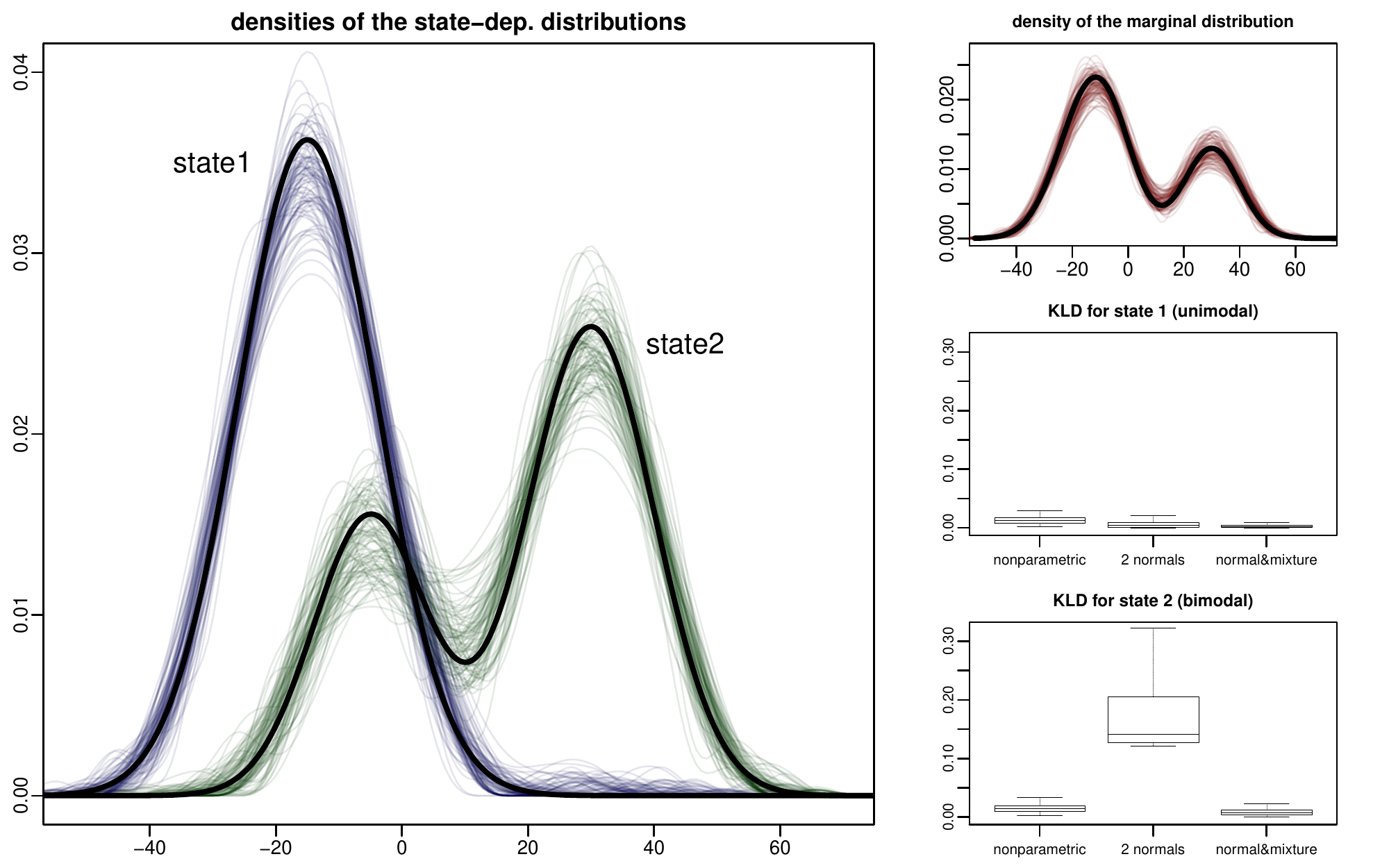}}
\end{center}
\caption{Simulation study: true and estimated densities of the state-dependent distributions estimated in the first 100 simulation runs (left plot, with true densities indicated by thick black lines), corresponding true and estimated densities of the marginal distribution (top right plot, with true density indicated by thick black line), and box plots of KLDs for unimodal state 1 (middle right plot) and for bimodal state 2 (bottom right plot).} \label{Fig:sim}
\end{figure}

To demonstrate the practicality of the suggested approach, we first conduct a simulation study. We consider a two-state HMM where the state-dependent distributions substantially overlap, with a unimodal conditional distribution in state 1 and a bimodal conditional distribution in state 2; see Figure \ref{Fig:sim} for an illustration. The states of the Markov chain were generated using the t.p.m.\
\begin{equation*}
\Gamma = \begin{pmatrix} 0.9 & 0.1 \\ 0.1 & 0.9 \end{pmatrix}  .
\end{equation*}
In practice, the chosen configuration would make it difficult to specify an adequate parametric HMM, since the marginal distribution gives no indication for the bimodality of state 2 (as can be seen in the top right panel of Figure \ref{Fig:sim}). It is also not clear a priori if a nonparametric approach can exploit the correlation over time in order to identify the smaller peak of the conditional bimodal distribution in state 2 (at $x=-5$), or if it fails to do so and wrongly allocates the corresponding observations to state 1.

For this model, we conducted 500 simulation runs, with $T=800$ observations being generated in each run. In each run, the nonparametric approach was applied, maximizing (\ref{plik}) using the optimizer \texttt{nlm} in R; corresponding code is given in Web Appendix B. For the cross-validation, we selected the grid of potential smoothing parameter vectors, $\boldsymbol{\Lambda}$, such that as possible values for each of the state-specific smoothing parameters the values $256$, $512$, $1024$, $2048$, $4096$, $8192$ and $16384$ were considered. We used $C=10$ cross-validation partitions in each simulation run to select the smoothing parameter vector from $\boldsymbol{\Lambda}$. We further set $K=15$, hence using 31 B-spline basis densities in the estimation.

The sample mean estimates of the transition probabilities $\gamma_{11}$ and $\gamma_{22}$ were $0.907$ (Monte Carlo standard deviation of estimates: $0.018$; average of standard errors obtained in each run via parametric bootstrap: $0.018$; coverage of bootstrap 95\% confidence intervals obtained in each run: $92.6\%$) and $0.907$ ($0.018$; $0.019$; $95.0\%$), respectively. The estimated state-dependent distributions from the first 100 simulation runs are visualized in the left panel of Figure \ref{Fig:sim}. All fits are fairly reasonable, although as expected the peaks are slightly underestimated, on average, while the troughs are slightly overestimated. The same pattern can be seen for the marginal distribution, displayed in the top right panel in Figure \ref{Fig:sim}. 

We further calculated the Kullback-Leibler divergence (KLD) of the estimated densities from those in the true model, for both state-dependent distributions and in each simulation run. To have a benchmark, we also calculated the corresponding KLDs of densities estimated using either of two parametric HMMs: 1) an incorrect parametric model, assuming normal state-dependent distributions (which one may visually deduce from a histogram of the observations to be appropriate), and 2) the correct parametric model, involving a normal distribution for state 1 and a mixture of two normal distributions in state 2. Unsurprisingly, the correct parametric model performed best in terms of the KLD. 
The nonparametric approach yielded the highest average KLD for the conditional distribution in state 1, due to the oversmoothing close to the peak. For the state-dependent distribution in state 2, our nonparametric approach yielded an average KLD of $0.016$, which is slightly higher than the corresponding average KLD obtained using the correct parametric specification ($0.010$), whereas for the incorrectly specified parametric model the corresponding average KLD was obtained as $0.290$, indicating the expected much poorer fit. For the correctly specified parametric model, the sample mean estimates of the transition probabilities $\gamma_{11}$ and $\gamma_{22}$ were obtained as $0.898$ (Monte Carlo standard deviation of estimates: $0.018$; average of standard errors obtained in each run via parametric bootstrap: $0.019$; coverage of bootstrap 95\% confidence intervals obtained in each run: $95.6\%$) and $0.899$ ($0.020$; $0.021$; $94.2\%$), respectively, while for the incorrectly specified parametric model the corresponding mean estimates were obtained as $0.866$ ($0.019$; $0.040$; $63.6\%$) and $0.821$ ($0.024$; $0.093$; $45.6\%$), respectively. The incorrectly specified model thus led to erroneous inference on the state process, in particular substantially underestimating the persistence in state 2. Overall, in this simulation study the nonparametric approach performed only slightly worse than the correct parametric specification --- unlike the latter leading to a small bias in the estimates of the transition probabilities, and resulting in slightly less accurately estimated state-dependent distributions  --- and much better than the incorrect parametric specification which one may naively choose based on a histogram of the data or another form of visual inspection.

In each simulation run, we additionally fitted one- and three-state HMMs with nonparametrically modeled state-dependent distributions, in order to illustrate that the cross-validation technique can also be used to select the number of states. In each run, the model selection was based on a comparison of the out-of-sample log-likelihood scores on 10 random validation samples, obtained for the different models fitted to the corresponding 10 calibration samples. This is essentially the multifold cross-validation procedure considered in \citet{cel08}, only that here we obtain estimates via direct maximization of the likelihood rather than using the expectation-maximization algorithm. 
In our simulations, this model selection exercise led to a correct identification of the two-state model in 459 out of 500 cases ($91.8\%$), with the three-state model being selected in the other 41 cases. 

To investigate the estimation performance under different conditions, we experimented with several further model formulations. In particular, we considered a) different levels of correlation as induced by the 2-state Markov chain (by varying the diagonal entries in the t.p.m.) and b) different levels of overlap of the two state-dependent distributions (by shifting one of the two). In all those scenarios where there was a reasonable level of correlation induced by the Markov chain (roughly, for both diagonal entries either $>0.75$ or $<0.25$), the estimation worked well. The estimation performance improved with diagonal entries in the t.p.m.\ approaching either 1 or 0 (which leads to an increased correlation). This intuitively makes sense, since the stronger the correlation, the clearer becomes the pattern and hence the easier it is for the model to allocate observations to states. Similarly, the estimation performance improved as the overlap of the state-dependent distributions was reduced, again due to the pattern becoming clearer.


\section{Modeling vertical speeds of a diving whale}
\label{appl}

\subsection{The data}

We consider a 48-hour time series of absolute depth displacements by a single adult female Blainville's beaked whale in Hawaii, tagged with a Mk9 time-depth recorder (Wildlife Computers, Redmond, WA, USA) and previously described by \citet{bai08}. 
This species performs deeper foraging dives and shallower non-foraging dives, with higher vertical speeds during deep dives, especially during descents \citep{bai08}. 

The original data give depth measurements once per second, but for the purpose of illustrating our methods it will be expedient to model depth displacements on a coarser temporal scale, with one observation per minute. We consider absolute values of depth displacements, hence focusing on speed and ignoring the direction. For modeling purposes, we take the logarithms of those values as it is more convenient to model a distribution whose support is the real line. 
Every observation thus gives the logarithm of the absolute vertical displacement of the whale over the previous minute, which is an indicator for the whale's vertical speed in that time period. The resulting time series to be modeled, comprising 2880 observations, is illustrated in the top panel in Figure \ref{obs2}, alongside a histogram of the observations and the sample autocorrelation function for the observed series (bottom left and bottom right panel, respectively). The multimodality depicted in the histogram is a consequence of the whale occupying different behavioral states at different times, and this pattern, together with the strong autocorrelation, motivates the use of dependent mixtures such as HMMs for these data.

\subsection{Analysis using nonparametric HMMs}

The histogram motivates the use of a nonparametric approach for modeling the state-dependent distributions of such an HMM, since it is not evident what would be a suitable parametric family. The different characteristics of the behavioral states of the whale, most notably ``close to the surface'', ``on the ascent/descent'' and ``at the bottom of a dive'', motivate the use of at least three states in an HMM for this time series. From the biological point of view, the results we obtained when fitting a 4-state nonparametric HMM did not offer further relevant insights. 
We also note that for models with more than 4 states, the numerical search routine became very sensitive to the choice of initial values used in the maximization, so that it was difficult to identify the maximum penalized likelihood estimate. Thus, since our objective here is to illustrate the nonparametric modeling approach, we restrict ourselves to the consideration of a 3-state model.  

\begin{figure}[!htb]
\begin{center}
{\includegraphics*[width=0.95\textwidth]{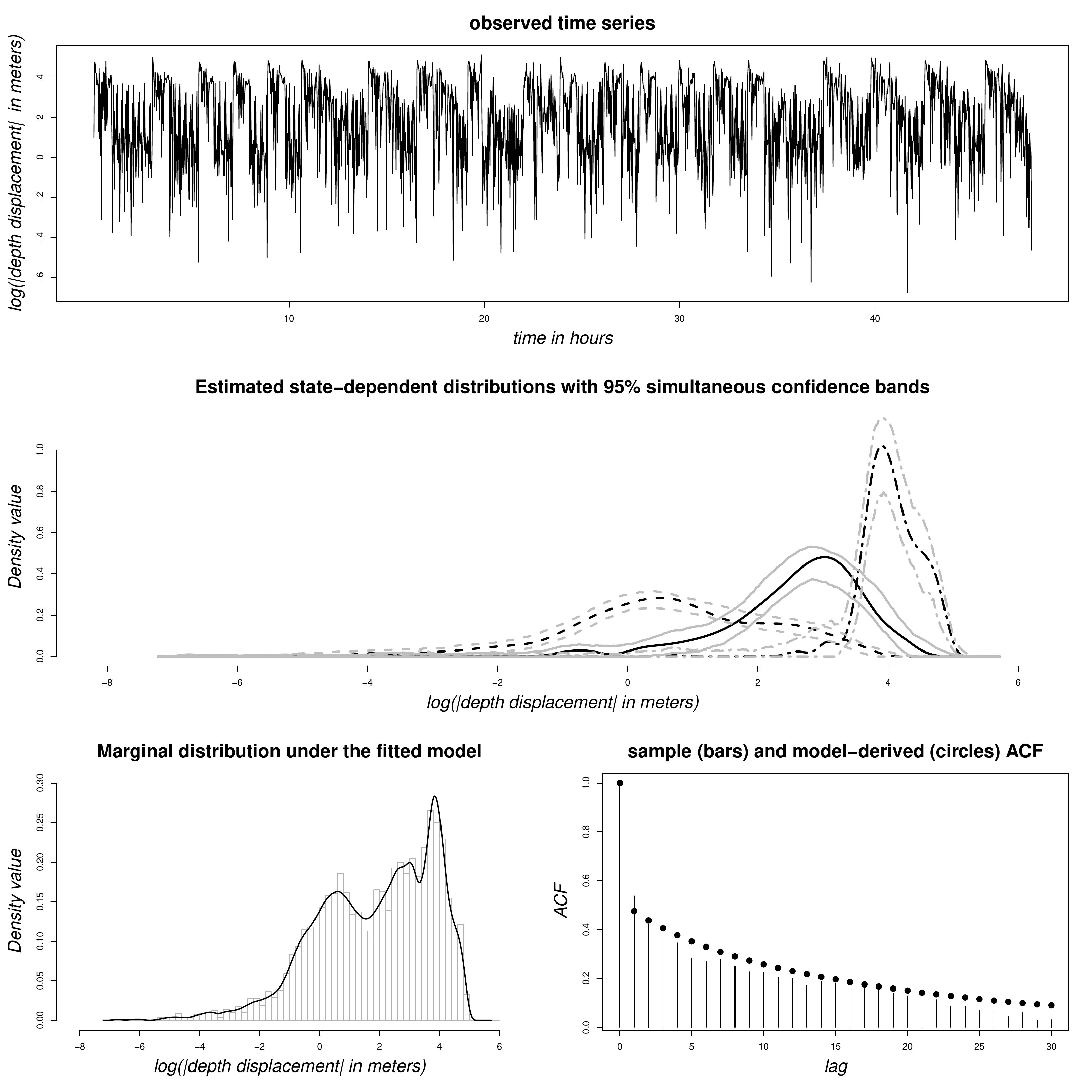}}
\end{center}
\caption{The top plot shows the time series that is modeled. The plot in the middle shows the estimated state-dependent distributions (weighted with their proportion in the mixture according to the stationary distribution of the Markov chain, and together with 95\% simultaneous confidence bands). The bottom left plot shows the corresponding marginal distribution (solid line), together with a histogram of the observations (grey bars). The bottom right plot gives the sample autocorrelation function (vertical bars) and the model-derived autocorrelation function (black circles). } \label{obs2}
\end{figure}

The 3-state (stationary) nonparametric HMM was fitted to the described series via maximum penalized likelihood estimation, maximizing (\ref{plik}) numerically.  
As smoothing parameter vector we selected $\boldsymbol{\lambda}=(65536,8192,32)$ via cross-validation as described in Section \ref{cv} (see Appendix A for more details). We used 51 standardized B-splines in the estimation of the state-dependent distributions, i.e., $K=25$ in (\ref{lincom2}). On an i7 CPU, at 2.7 GHz and with 4 GB RAM, the parameter estimation took about 20 minutes using R, which could be reduced to about 2 minutes by writing the forward algorithm in C++. To minimize the risk of missing the global maximum of the likelihood, 500 randomly chosen sets of initial values were tried. The t.p.m.\ was estimated as
\begin{equation*}
 \widehat{\boldsymbol{\Gamma}}=\begin{pmatrix}
0.975 \; {\it (0.965,0.983)} & 0.007 \; {\it (0.001,0.015)} & 0.018 \; {\it (0.010,0.027)} \\
0.017 \; {\it (0.005,0.033)} & 0.893 \; {\it (0.871,0.926)} & 0.090 \; {\it (0.059,0.113)} \\
0.038 \; {\it (0.020,0.056)} & 0.111 \; {\it (0.074,0.138)} & 0.851 \; {\it (0.821,0.890)}
\end{pmatrix} ,
\end{equation*}
with the 95\% confidence intervals given in the brackets obtained using a parametric bootstrap (500 samples). Figure \ref{obs2} displays 1) the time series that was modeled, 2) the state-dependent distributions of the fitted model, together with 95\% simultaneous confidence bands (obtained as described in Section \ref{se} based on the 500 bootstrap samples), 3) the marginal distribution of $X_t$ according to the fitted model, together with a histogram of the observations, and 4) the sample autocorrelation function alongside the model-derived autocorrelation function. An illustration of how the fitted state-dependent distributions are built from the B-spline basis densities is given in Appendix A (Figure A1). 

To assess the goodness-of-fit of the fitted model, we calculated the one-step-ahead forecast pseudo-residuals \citep{zuc09},
$$r_t=\Phi^{-1}\bigl( F(x_t \mid x_{t-1}, x_{t-2}, \ldots , x_1) \bigr), \quad t=1,\ldots,T,$$
where $\Phi$ is the cumulative distribution function (c.d.f.) of the standard normal distribution, and $F(x_t \mid x_{t-1}, x_{t-2}, \ldots , x_1)$ is the c.d.f.\ of the one-step-ahead forecast distribution at time $t-1$, under the fitted model, evaluated at $x_{t}$:
\begin{equation*}
F(x_{t} \mid x_{t-1},x_{t-2}, \ldots , x_1)   =  \sum_{i=1}^N \zeta_{t,i} \Pr (X_t \leq x_{t} \mid S_t = i)
\nonumber                                    =  \sum_{i=1}^N \zeta_{t,i} \int_{-\infty}^{x_t} f_i \left( z \right) dz  \, .
\end{equation*}
Here $\zeta_{1,i}=\delta_i$ and for $t=2,\ldots,T$, $\zeta_{t,i}$ is the $i$th entry of the vector $\boldsymbol{\alpha}_{t-1} \boldsymbol{\Gamma}/(\boldsymbol{\alpha}_{t-1}\mathbf{1})$, where 
$\boldsymbol{\alpha}_{t-1}=\boldsymbol{\delta} \mathbf{Q}(x_1) \boldsymbol{\Gamma} \mathbf{Q}(x_2) \ldots \boldsymbol{\Gamma} \mathbf{Q}(x_{t-1})$ is the {\em forward variable}. If the model is correct, then the pseudo-residuals are distributed standard normal. A quantile-quantile plot of the residuals against the standard normal, and the sample autocorrelation function of the series of residuals, are given in Appendix A (Figure A3). 
The plots indicate a very good fit and only a minor correlation in the residuals over time. Applying a Jarque--Bera test to the pseudo-residuals yields a p-value of 0.3, such that the null hypothesis of normality cannot be rejected at the 5\% level. The model hence fits the data well.

For the fitted model, we applied the Viterbi algorithm to find the sequence of states $s_1^*,\ldots,s_T^*$ that is most likely to have given rise to the data (\citealp{zuc09}).
To facilitate interpretation, we compared the decoded states to the actual positions of the whale in the water column (which were recorded, but not modeled here);
an illustration is given in Appendix A (Figure A2). We find that state 1 of the fitted HMM, which is associated with the smallest absolute depth displacements, captures the whale's vertical speeds close to the surface and on very shallow dives (to depths $<100$ meters), with the shallow dives causing the second mode in the fitted density (at values slightly higher than 2). This state is occupied about 52\% of the time according to the stationary distribution of the fitted Markov chain. State 2, which involves moderate absolute depth displacements, is occupied about 26\% of the time and is associated with likely foraging periods at the bottoms of deep dives. 
State 3 implies the highest absolute depth displacements, is occupied about 22\% of the time and only on deep dives, and is occasionally interspersed with state 2 at the bottoms of those dives due to slower movement related to foraging activity. 
We note that any biological interpretations of this model have to be made cautiously. In particular, the HMM states are not to be confounded with behavioral states of the animal, as they merely summarize vertical diving speeds into three categories, and the speeds can be similar for distinct behaviors. However, it can nevertheless be stated that the features implied by the fitted nonparametric HMM are consistent with previous research on the species \citep{bai08}.
Moreover, this exploratory analysis demonstrates the potential of these models as tools for example for objective identification and characterisation of foraging periods, which is notoriously challenging with time-depth recorder data \citep{hoo01}. 

\subsection{Comparison with conventional parametric HMMs}

In order to have a benchmark for the above results, and to illustrate the potential usefulness of the nonparametric approach in this type of application (and in fact also more generally), we additionally considered conventional parametric HMMs. For the given time series, HMMs with normal
state-dependent distributions constitute an obvious (and plausible) choice of a parametric family of models. Thus, we present the results of fitting such models to the described data, acknowledging that this is only one of dozens of plausible parametric HMM formulations that could be considered -- a flexibility which is both a blessing and a curse when dealing with HMMs, as the model formulation step is in general by no means straightforward. 

The results of fitting HMMs with normal state-dependent distributions and 3--10 states, and some graphic illustrations of the fitted models, are provided in Appendix A. From these models, the Akaike Information Criterion and the Bayesian Information Criterion select models with 10 and 7 states, respectively. (We did not consider models with even higher numbers of states because of increasing numerical instability.) The need for these high numbers of states is confirmed by goodness-of-fit analyses of the fitted models, where normality of the pseudo-residuals is rejected by a Jarque--Bera test, at the 5\% level, for all models with less than 7 states. For the 7-state parametric HMM, the pseudo-residuals indicate a goodness-of-fit similar to that of the 3-state nonparametric HMM. Figures A4 and A6 in the Web Appendix illustrate that a crucial problem with the considered parametric formulation is the lack of flexibility of the state-dependent distributions, with the consequence that the marginal distribution can not be captured adequately with small numbers of states. Indeed, for the 7-state model --- which based on the information criteria and the goodness-of-fit test seems to be a plausible model --- it is difficult to come up with biologically meaningful interpretations of the states, with the results indicating that some of the HMM states may in fact be lumped together to form a single behavioral state (cf.\ Figure A8, states 3 and 5). 



\subsection{Considerations related to biological inference}

Overall, there are a number of reasons why biologists may prefer working with the nonparametric model. The ability to summarise the data accurately without using a large number of states facilitates interpretation, with the nonparametric approach resulting in greater persistence within states, 
so that intuitive association of the states with broader behavior is more straightforward. The reduced tendency of the nonparametric model to allocate additional states simply to accurately capture the marginal distribution should hence allow more efficient characterization of behavioral responses to disturbance.  For example, beaked whale responses to military sonar sounds can include an initial increase in vertical velocity as the animal dives deeply, fewer quick vertical excursions during the bottom phase of the dive, and long, slow ascents \citep{tya11}. 
Our model could help determine whether these changes are best described in terms of ``response'' states (outside the realm of usual vertical displacement patterns), adjustments to peaks and variability of the marginal distribution, or unusual inter-state transitions. 
The ability to distinguish truly novel states from more subtle alterations of an (already complex) behavior regime is key to assessing the biological and management consequences of behavioral responses to disturbance. A model that accurately summarises the data yet is parsimonious is ideally suited for these purposes, since higher numbers of parameters in the t.p.m.\ render it much less practical to incorporate for example covariates, but also random effects, in the state process.



\section{Discussion}
\label{discuss}

Exploiting the strengths of the HMM machinery and of penalized B-splines, we have developed a relatively simple yet powerful and flexible likelihood-based framework for a nonparametric estimation of the state-dependent distributions in an HMM. The framework lends itself to comprehensive inferential analyses, including uncertainty quantification, model checking and state decoding. In our view, the choice of the smoothing parameter, local maxima of the likelihood and uncertainty quantification constitute the most challenging issues with the approach, and future work should explore ways to enhance the feasibility in these regards. For smoothing parameter selection, as an alternative to cross-validation, one could consider a leverage-based, approximate leave-one-out cross-validation which yields an AIC-type criterion. Regarding local maxima, estimation via the EM algorithm could potentially be more robust \citep{bul08}, but is technically more challenging and likely to be slower. Uncertainty quantification is routinely achieved in the Bayesian approach of \citet{yau11} by studying the variability of the posterior samples, whereas we employed computationally expensive bootstrap techniques.

Applying the proposed nonparametric approach to a time series of beaked whale vertical displacement data, we have illustrated that it can produce a coherent, succinct summary of the data using a small number of states, sensibly partitioning the data into few velocity regimes that are easy to relate to broader behavior categories. The nonparametric approach essentially offers unlimited flexibility to capture the marginal distribution, irrespective of the number of states considered, which means that inference, particularly on the number of states, is based solely on the correlation structure of a given time series. In contrast, in our case study we have seen that conventional parametric HMMs can lead to high numbers of states being selected due to limited flexibility of the state-dependent distributions considered, leading to state processes that are overly complex relative to the actual correlation structure. This reveals a potentially substantial benefit of the nonparametric approach, as the disentanglement of the two main reasons for a poor fit of an HMM --- failure to capture the marginal distribution and failure to capture the correlation structure --- can be quite challenging using conventional methods. In practice, the nonparametric approach will often lead to models that are more parsimonious in terms of the number of states than parametric counterparts. In the context of measuring behavioral responses to acoustic disturbance, the approach promises enhanced ability to arrive at a nuanced characterization of different responses.  

In general, it will often be the case that multiple time series, for example associated with multiple individuals, are collected. In the given type of application, this will in fact often be necessary in order to adequately address biological questions of interest, e.g., regarding the effect of sonar exposure. Models for such longitudinal data typically account for potential variability between the different component series, by allowing some model parameters to be component-specific, while borrowing strength across component series in order to improve estimator precision. \citet{zuc08} give a useful overview of the different strategies for modeling heterogeneity of HMM component series. Regarding the state process of an HMM with nonparametrically modeled state-dependent distributions, the techniques provided by \citet{alt07} in her framework of mixed HMMs --- which allows for random effects, but also covariates that are specific to the component series --- and the discrete random effects approach suggested by \citet{mar09} are directly applicable. The latter type of approach is nowadays routinely used for example in capture-recapture models (see the mixture models discussed by \citealp{ple03}), and using our approach it is conceptually straightforward to implement this idea also in the state-dependent process. However, with such a model formulation the number of parameters to be estimated increases considerably, likely rendering this extension of our approach practically infeasible in many scenarios with longitudinal data. Future research could consider alternative extensions along the lines of additive mixed models where individual-specific random curves can be estimated (see, e.g., \citealp{sch14}). On the other hand, we anticipate that in some applications it will be perfectly reasonable to assume state-dependent distributions to be common to all component series, with heterogeneity modeled only in the state-switching behavior.

There are several other interesting modifications and extensions of our approach. First, it is straightforward to consider semiparametric versions where some of the state-dependent distributions are modeled nonparametrically and others taken from a parametric class of distributions (as in \citealp{dan12}). This can improve numerical stability and decrease the computational burden associated with the cross-validation. Second, the likelihood-based approach allows for the consideration of more involved dependence structures. In particular, semi-Markov state processes --- where the duration of a stay in any given state is explicitly modeled, rather than implicitly assumed to be geometrically distributed, as in standard HMMs --- can easily be accommodated within the suggested approach using the representation proposed by \citet{lan11}. The consideration of multivariate state-dependent distributions is also possible, in principle, using tensor products of univariate basis functions. Finally, the HMM forward algorithm and P-splines could be used in concert in order to consider nonparametric versions of Markov-switching regression models \citep{ham89}.



%



% ...


\end{spacing}
 
\newpage

\renewcommand{\thetable}{A\arabic{table}}
\setcounter{table}{0}

\renewcommand{\thefigure}{A\arabic{figure}}
\setcounter{figure}{0}

\renewcommand{\thesection}{A\arabic{figure}}
\setcounter{section}{0}

\section*{Appendix A}

\subsection*{Additional information, illustrations and goodness-of-fit analyses for the 3-state nonparametric HMM}

The cross-validation procedure for selecting an appropriate smoothing parameter vector was conducted as follows. We initially calculated the scores for the vectors (64,64,64), (128,128,128), (256,256,256), ..., (8192,8192,8192) and (16384,16384,16384), in each of $C=50$ cross-validation partitions using 90\% of the data points in the calibration stage. The cross-validation procedure selected $(1024,1024,1024)$ from these candidate smoothing parameter vectors. Running the iterative procedure described in Section 2.2 in the main manuscript, with the initial point $\boldsymbol{\lambda}_0^*=(1024,1024,1024)$, then led to the choice of $(65536,8192,32)$ after 14 iterations. All neighbouring smoothing parameter vectors on the considered grid had only slightly worse log-likelihood scores on the validation samples, which is not surprising given that  small changes in individual smoothing parameters lead to only minimal changes in the smoothness of the fitted densities.

\begin{figure}[!htb]
\begin{center}
\includegraphics[width=0.85\textwidth]{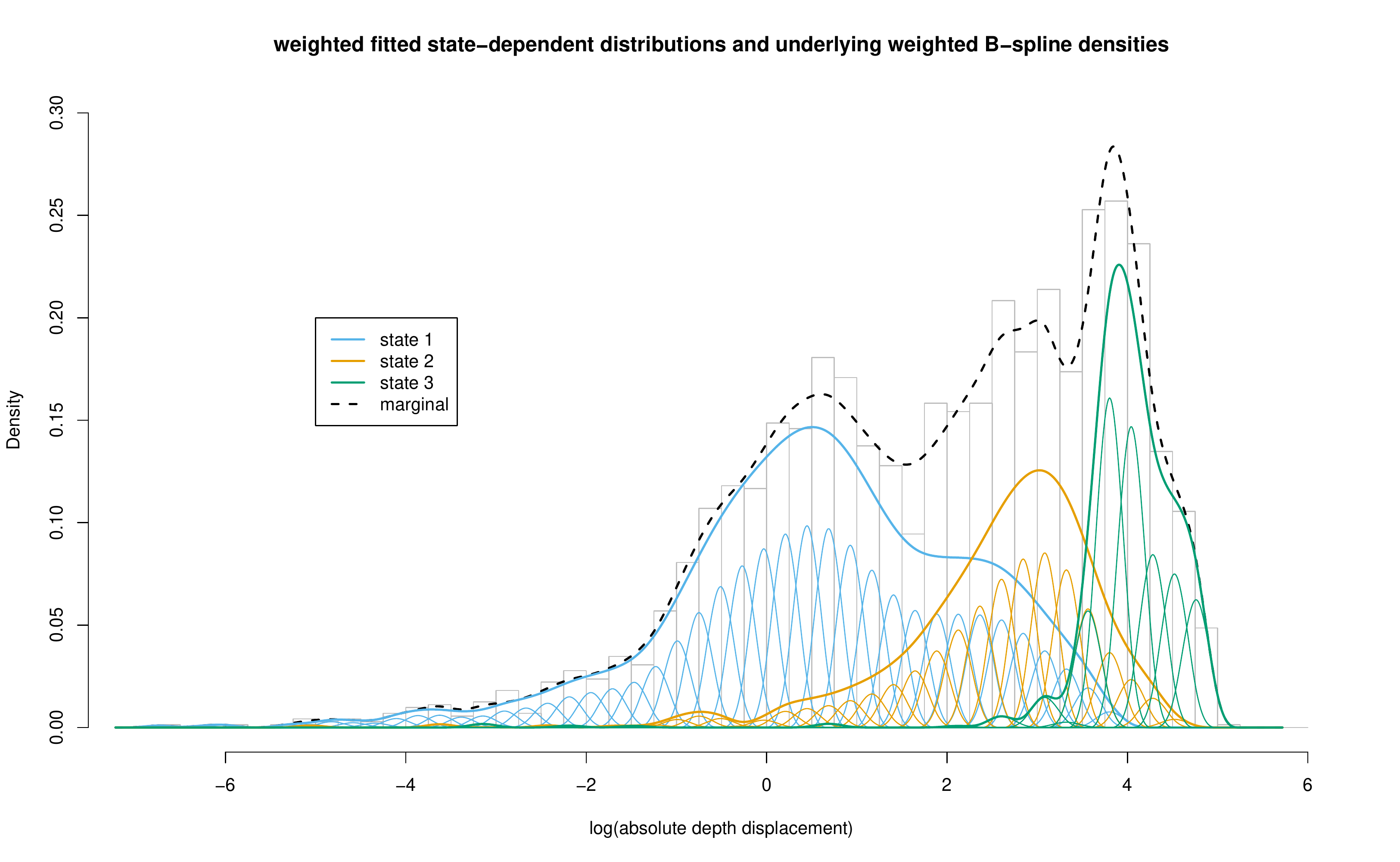}
\end{center}
\vspace{-2em}
\caption{\it Histogram of the observations, densities of the state-dependent distributions obtained in the real data example (each weighted with the corresponding proportion in the stationary distribution) with underlying weighted B-splines that generate these densities via a linear combination, and corresponding marginal distribution under the fitted model. } \label{bsplines}
\end{figure}

Figure \ref{bsplines} displays the densities of the state-dependent distributions estimated in the real data example, together with the corresponding B-spline basis densities that underlie the densities. The estimated densities were weighted with their proportion in the stationary distribution of the fitted Markov chain, and the displayed B-spline basis densities were weighted with the same proportion and with the estimated scalars that appear in the linear combination in Eq.\ (3) in the main manuscript.

\begin{figure}[!htb]
\begin{center}
\includegraphics[width=1\textwidth]{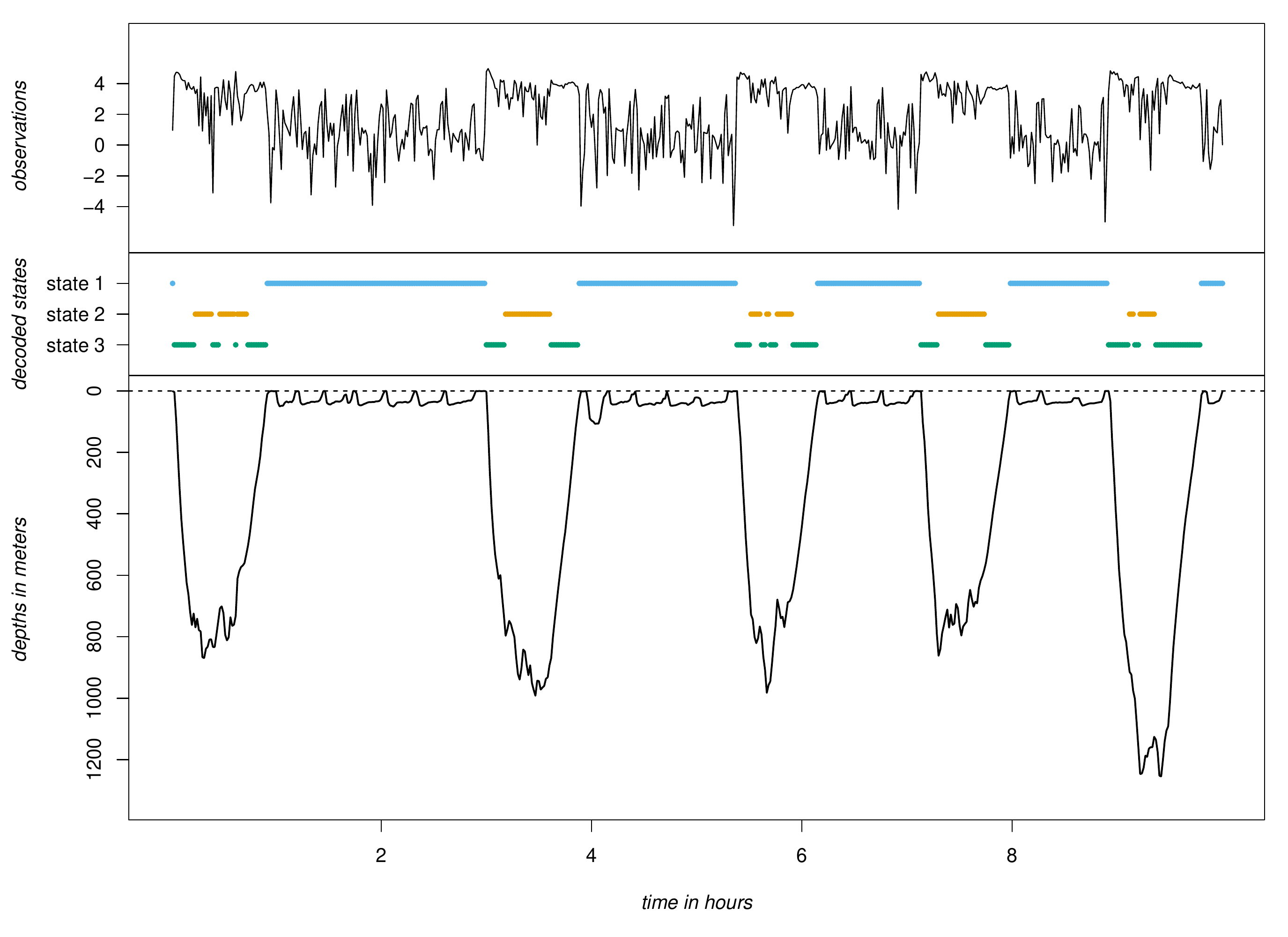}
\end{center}
\vspace{-2em}
\caption{\it Logarithms of absolute depth displacements (top panel), sequence of states that under the fitted 3-state model with nonparametric state-dependent distribtions is most likely to have given rise to the observations (middle panel), and actual depths, which were not modeled (bottom panel), for the first 10 hours of data.} \label{viterbi}
\end{figure}

\begin{figure}[!htb]
\begin{center}
\includegraphics[width=0.75\textwidth]{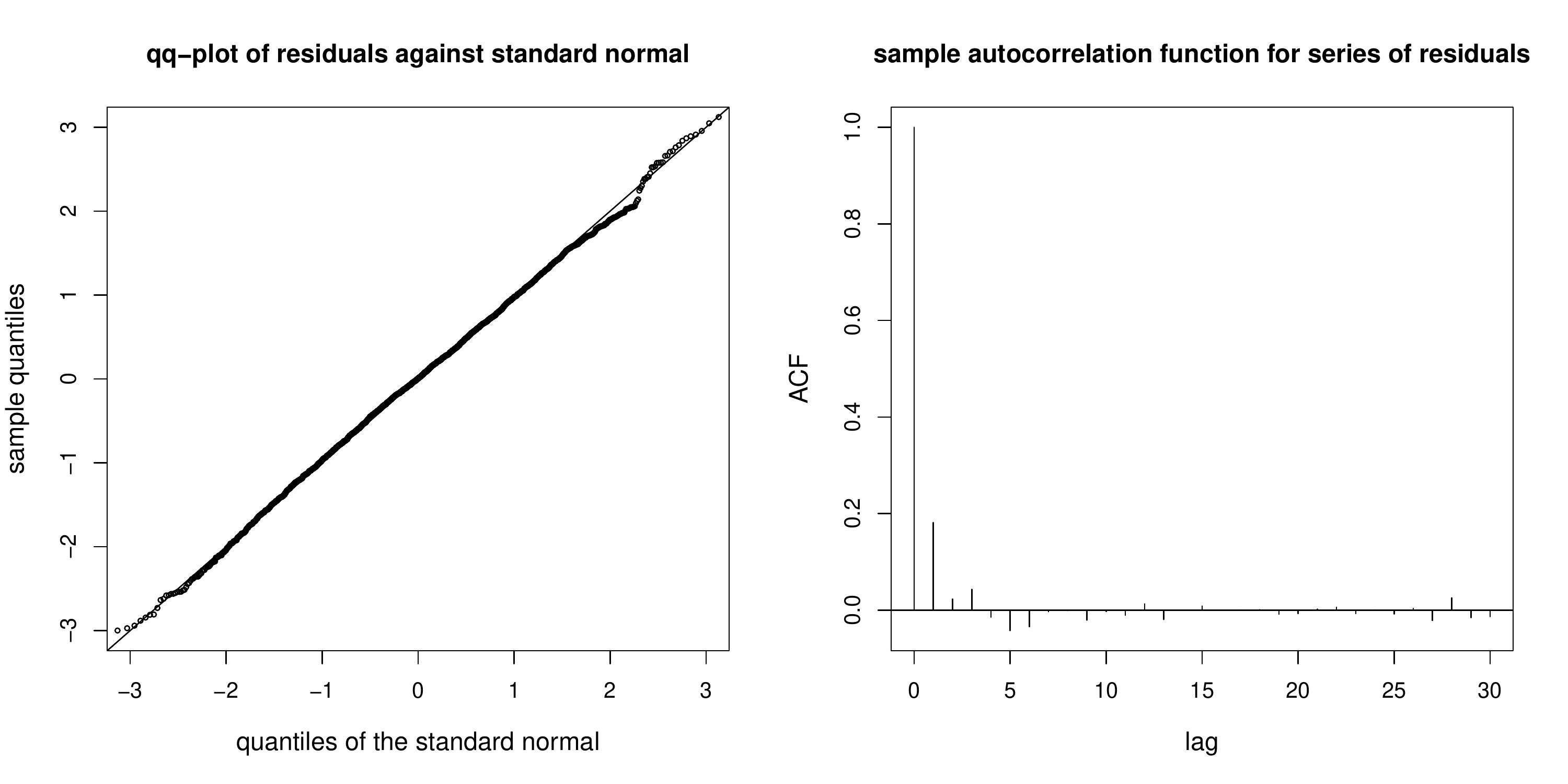}
\end{center}
\vspace{-1em}
\caption{\it Quantile-quantile plot of the one-step-ahead forecast pseudo-residuals obtained for the fitted nonparametric HMM against the standard normal distribution (left panel), and sample autocorrelation function of the series of residuals (right panel).} \label{qq}
\end{figure}

Figure \ref{viterbi} displays the decoded states underlying the first 10 hours of observations, i.e., $s_1^*,\ldots,s_{600}^*$. The figure also shows both the series modeled here and the actual depths observed during that time period (which were not modeled, and are shown only for the purpose of checking the biological meaningfulness of the fitted model). Figure \ref{qq} displays a quantile-quantile plot of the forecast pseudo-residuals against the standard normal distribution, and the sample autocorrelation function of the series of residuals.

\vspace{2em}

\newpage

\subsection*{Additional information, results, goodness-of-fit analyses and illustrations for conventional parametric HMMs}

Table \ref{resultspara} displays a summary of the results obtained when fitting conventional parametric HMMs with normal state-dependent distributions to the beaked whale data, including the log-likelihood values, the AIC values, the BIC values and the p-values of Jarque--Bera tests for normality applied to the models' pseudo-residuals. Models with 3-10 states were considered, and in each case 1000 different sets of random initial values were tried in the numerical maximization of the likelihood, in order to reduce the risk of missing the global maximum.

\begin{table}[!htb]
 \caption{\it Results of fitting HMMs with normal state-dependent distributions to the beaked whale data ($N$: number of states; $p$: number of model parameters; $\log \mathcal{L}$: maximum of the log-likelihood; AIC: Akaike Information Criterion; BIC: Bayesian Information Criterion; JB p-value: p-value of the Jarque--Bera test for normality applied to the pseudo-residuals.)}
\label{resultspara}
\begin{center}
\begin{tabular}{ccccccccccc}
\hline\\[-0.8em]
$N$      &    $p$   &     $\log \mathcal{L}$       &       AIC     &      BIC     &    JB p-value \\[0.2em]
\hline\\[-0.8em]
3        &     12   &     -4880.00                 &      9784.00  &    9855.59   &    0.000  \\
4        &     20   &     -4729.08                 &      9498.16  &    9617.47   &    0.000  \\
5        &     30   &     -4670.15                 &      9400.30  &    9579.27   &    0.002  \\
6        &     42   &     -4605.44                 &      9294.88  &    9545.43   &    0.016  \\
7        &     56   &     -4548.02                 &      9208.04  &    {\bf 9542.11}   &    0.261  \\
8        &     72   &     -4492.57                 &      9129.15  &    9558.67   &    0.310  \\
9        &     90   &     -4455.48                 &      9090.98  &    9627.87   &    0.475  \\
10       &    110   &     -4422.26                 &      {\bf 9064.53}  &    9720.74   &   0.429  \\
\hline \\[1em]
\end{tabular}
\end{center}
\end{table}

\begin{figure}[!htb]
\begin{center}
\includegraphics[width=1\textwidth]{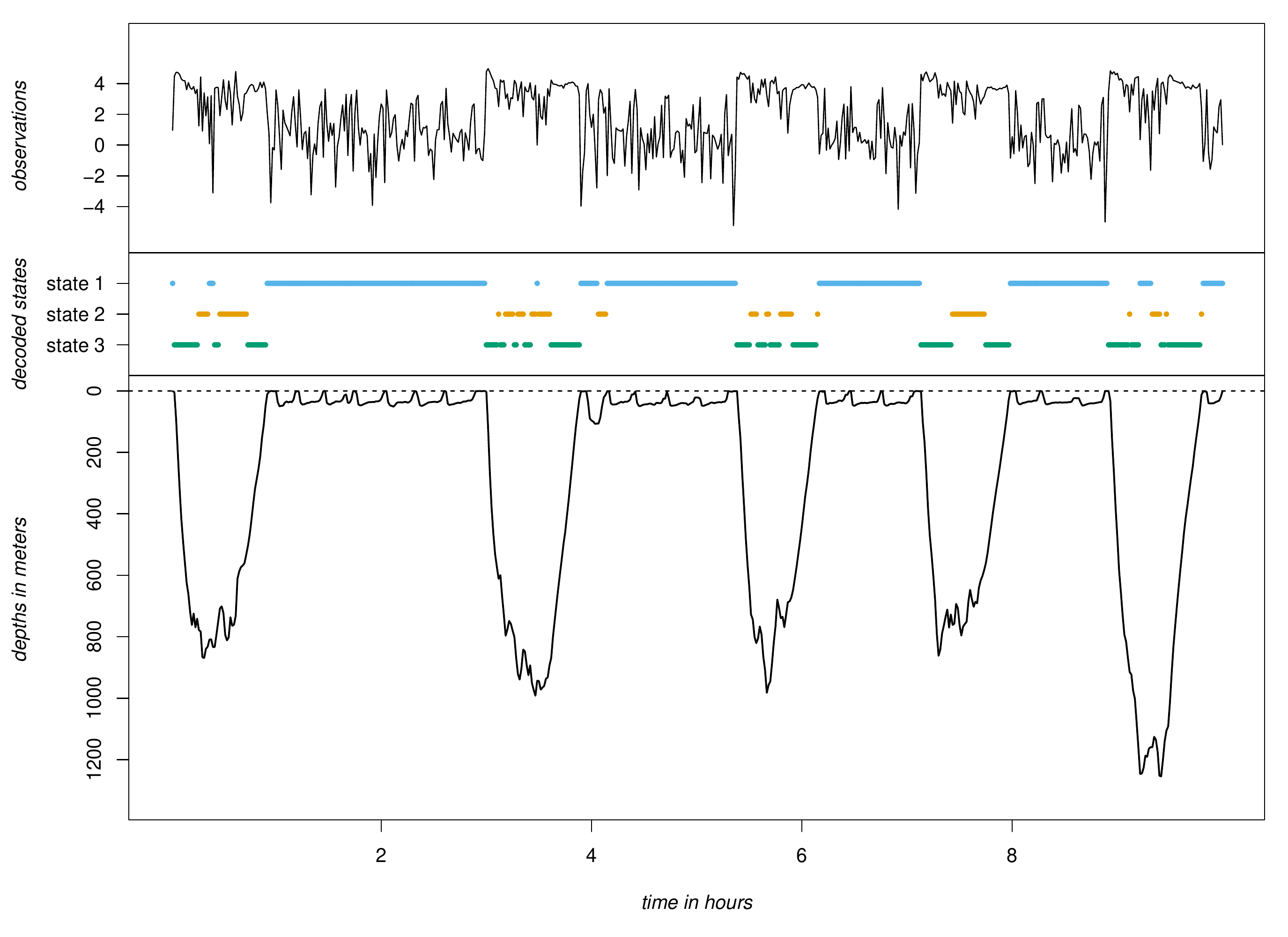}
\end{center}
\vspace{-2em}
\caption{\it Logarithms of absolute depth displacements (top panel), sequence of states that under the fitted 3-state model with normal state-dependent distributions is most likely to have given rise to the observations (middle panel), and actual depths, which were not modeled (bottom panel), for the first 10 hours of data.} \label{viterbipara3}
\end{figure}

The information criteria suggest that, from this parametric family, models with about 7-10 states are most appropriate. This is corroborated by the analysis of the associated pseudo-residuals, for which a Jarque--Bera test rejects normality at the 5\% level for all models with less than 7 states. The reason for the latter is the failure of the models with small numbers of states to capture the marginal distribution, due to insufficient flexibility of the state-dependent distribution applied (cf.\ Figures \ref{3state} and  \ref{qq3} below). The pattern picked up by the parametric 3-state model is the same as that picked up by the nonparametric 3-state model, with the three states corresponding to ``close to the surface'', ``on the ascent/descent'' and ``at the bottom of a dive'', respectively (cf.\ Figure \ref{viterbipara3}). However, biologically meaningful and relevant nuances in the data, such as for example the additional modes identified by the nonparametric model in states 1 (at values around 2.5) and 3 (at values around 4.8), are not captured by the parametric 3-state model due to its inflexibility.

\begin{figure}[!htb]
\begin{center}
\includegraphics[width=1\textwidth]{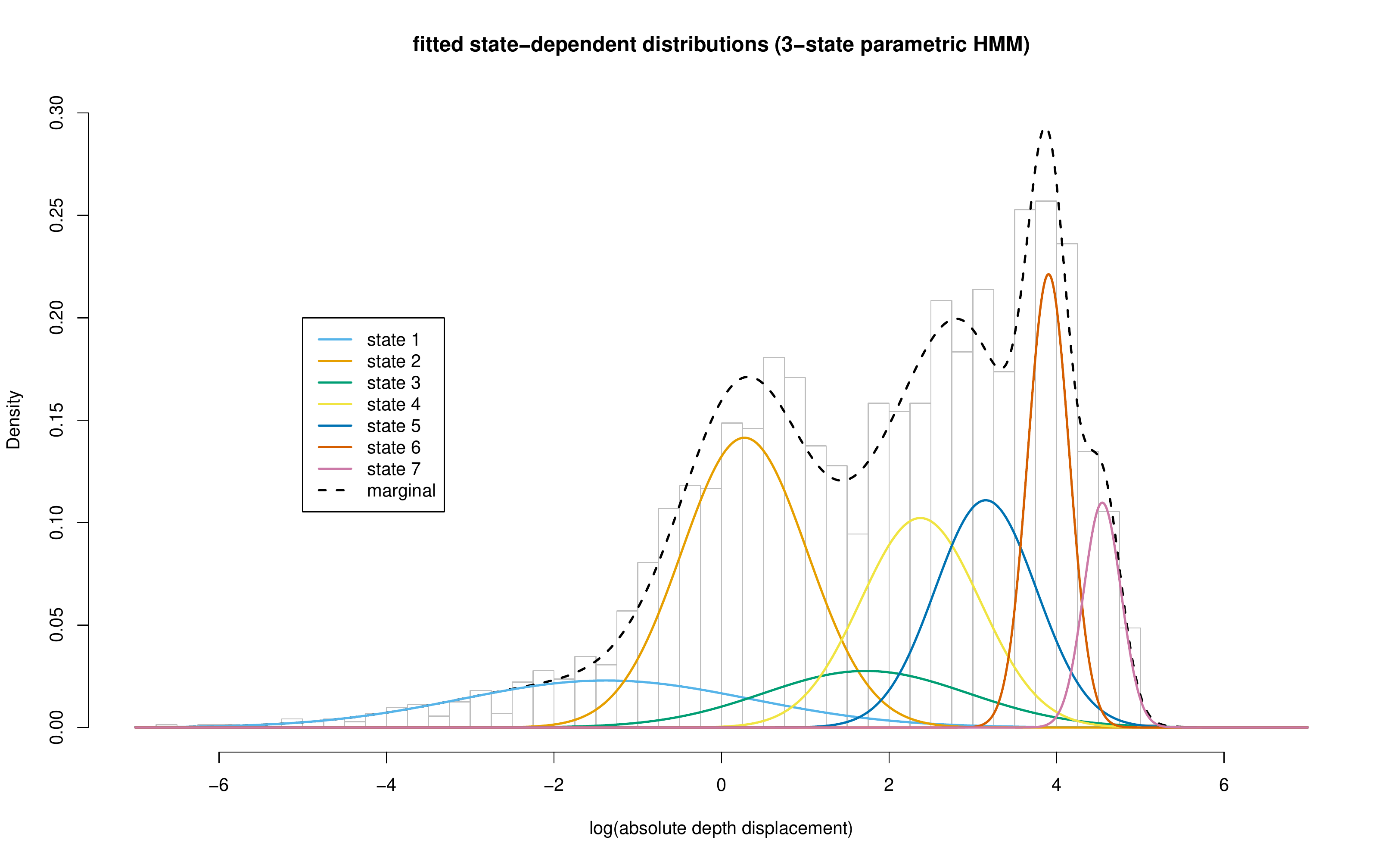}
\end{center}
\vspace{-2em}
\caption{\it Histogram of the observations, densities of the state-dependent distributions of the fitted 7-state parametric HMM (each weighted with the corresponding proportion in the stationary distribution), and corresponding marginal distribution under the fitted model.} \label{7state}
\end{figure}

\begin{figure}[!htb]
\begin{center}
\includegraphics[width=1\textwidth]{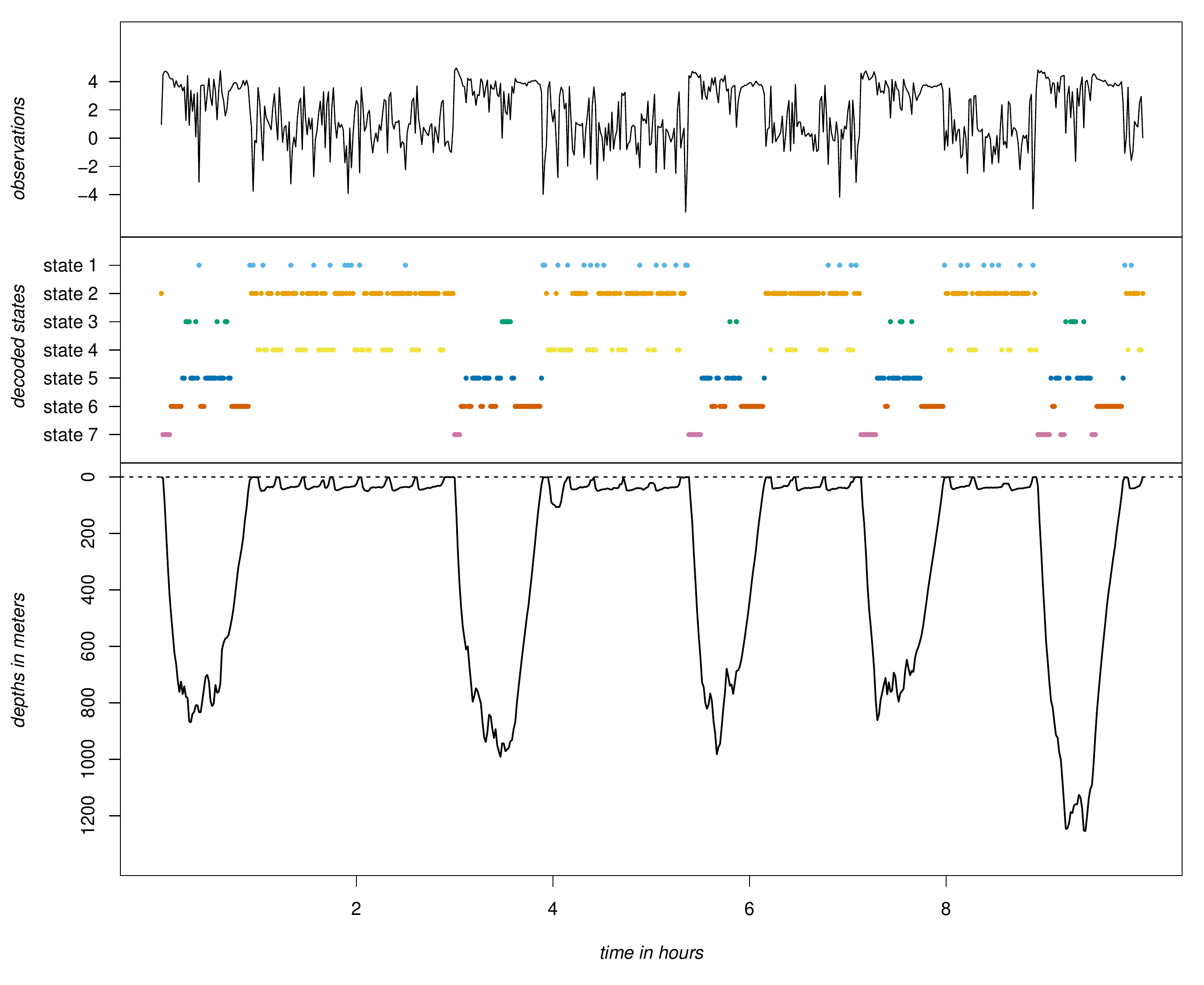}
\end{center}
\vspace{-2em}
\caption{\it Logarithms of absolute depth displacements (top panel), sequence of states that under the fitted 7-state model with normal state-dependent distributions is most likely to have given rise to the observations (middle panel), and actual depths, which were not modeled (bottom panel), for the first 10 hours of data.} \label{viterbipara7}
\end{figure}

\begin{figure}[!htb]
\begin{center}
\includegraphics[width=0.75\textwidth]{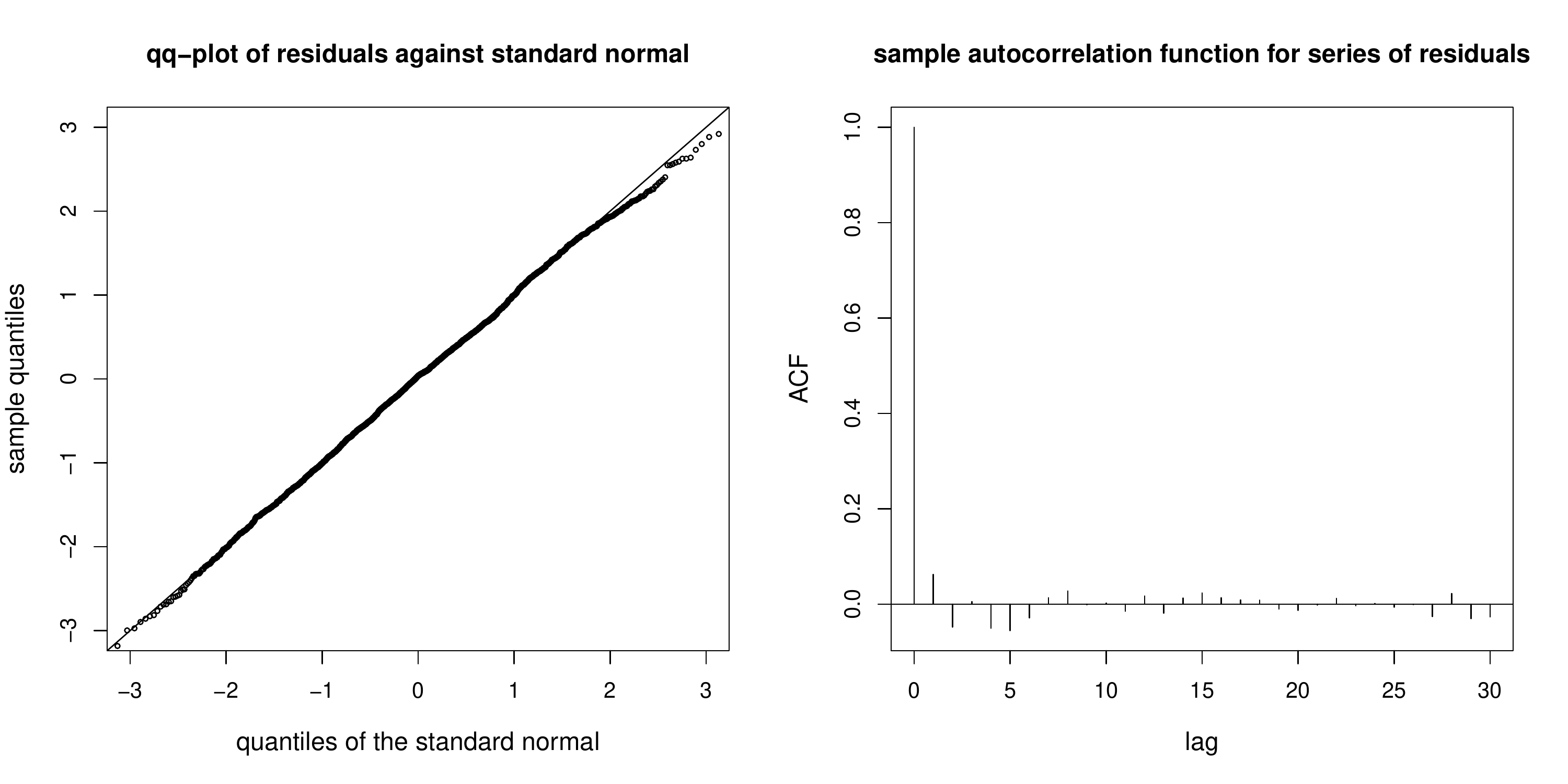}
\end{center}
\vspace{-1em}
\caption{\it Quantile-quantile plot of the one-step-ahead forecast pseudo-residuals obtained for the fitted parametric 7-state model against the standard normal distribution (left panel), and sample autocorrelation function of the series of residuals (right panel).} \label{qq7}
\end{figure}

To give an illustration of the consequences of increasing the number of states, Figures \ref{7state}, \ref{viterbipara7} and \ref{qq7} illustrate the state-dependent distributions, Viterbi-decoded states and analyses of the pseudo-residuals, respectively, for the fitted 7-state model. The marginal distribution is captured well by this more complex model, and the pseudo-residuals indicate a good fit. However, it is difficult to assign biologically meaningful interpretations to the states. For example, Figure A8 suggests that states 3 and 5 should probably be lumped together to form a single state, as they are associated with what seems to be a single behavioural mode of the animal. The distinction of this mode into two HMM states seems to be an artefact resulting from the limited flexibility of the state-dependent distributions applied.

\begin{figure}[!htb]
\begin{center}
\includegraphics[width=1\textwidth]{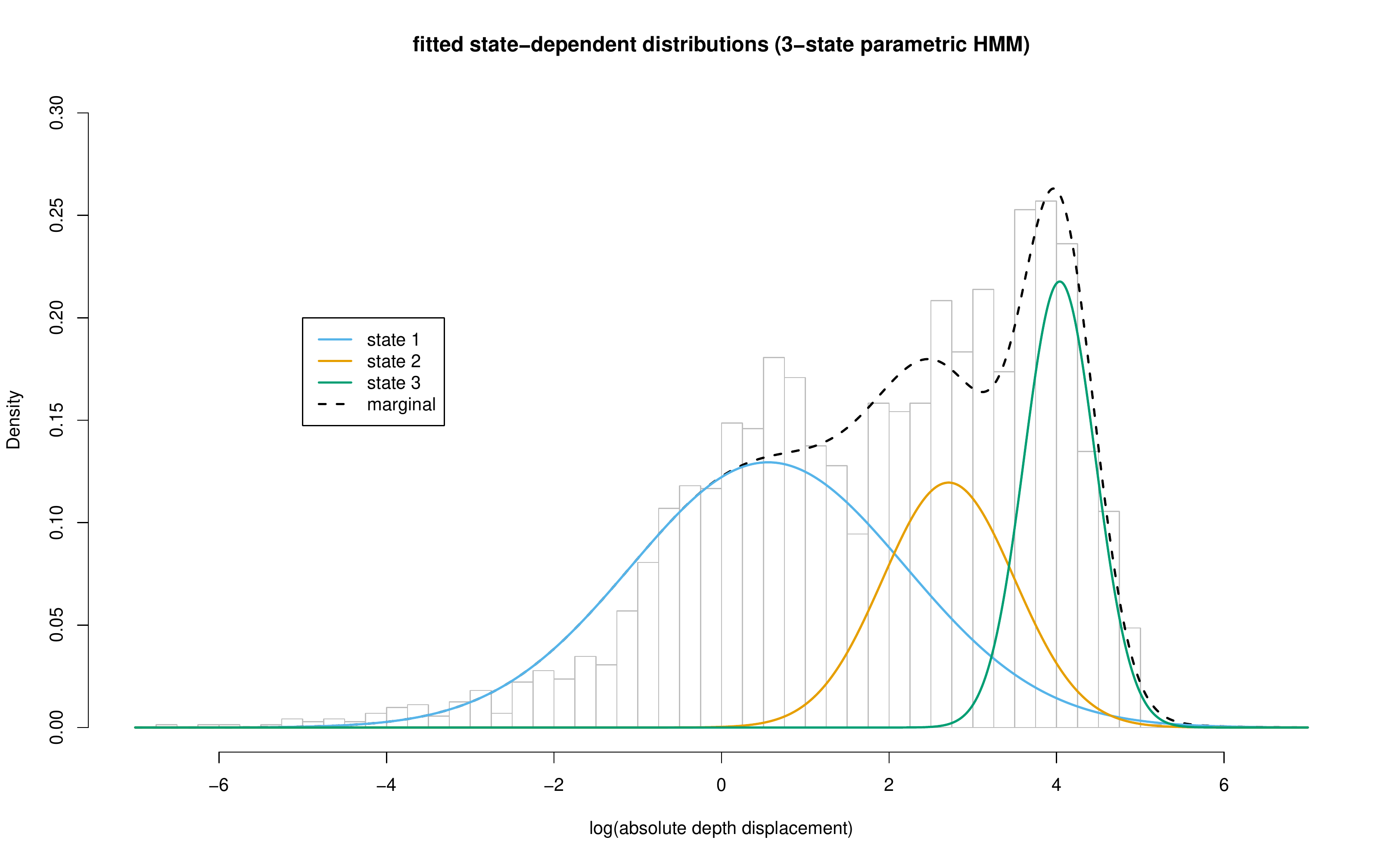}
\end{center}
\vspace{-2em}
\caption{\it Histogram of the observations, densities of the state-dependent distributions of the fitted 3-state parametric HMM (each weighted with the corresponding proportion in the stationary distribution), and corresponding marginal distribution under the fitted model. } \label{3state}
\end{figure}


\begin{figure}[!htb]
\begin{center}
\includegraphics[width=0.75\textwidth]{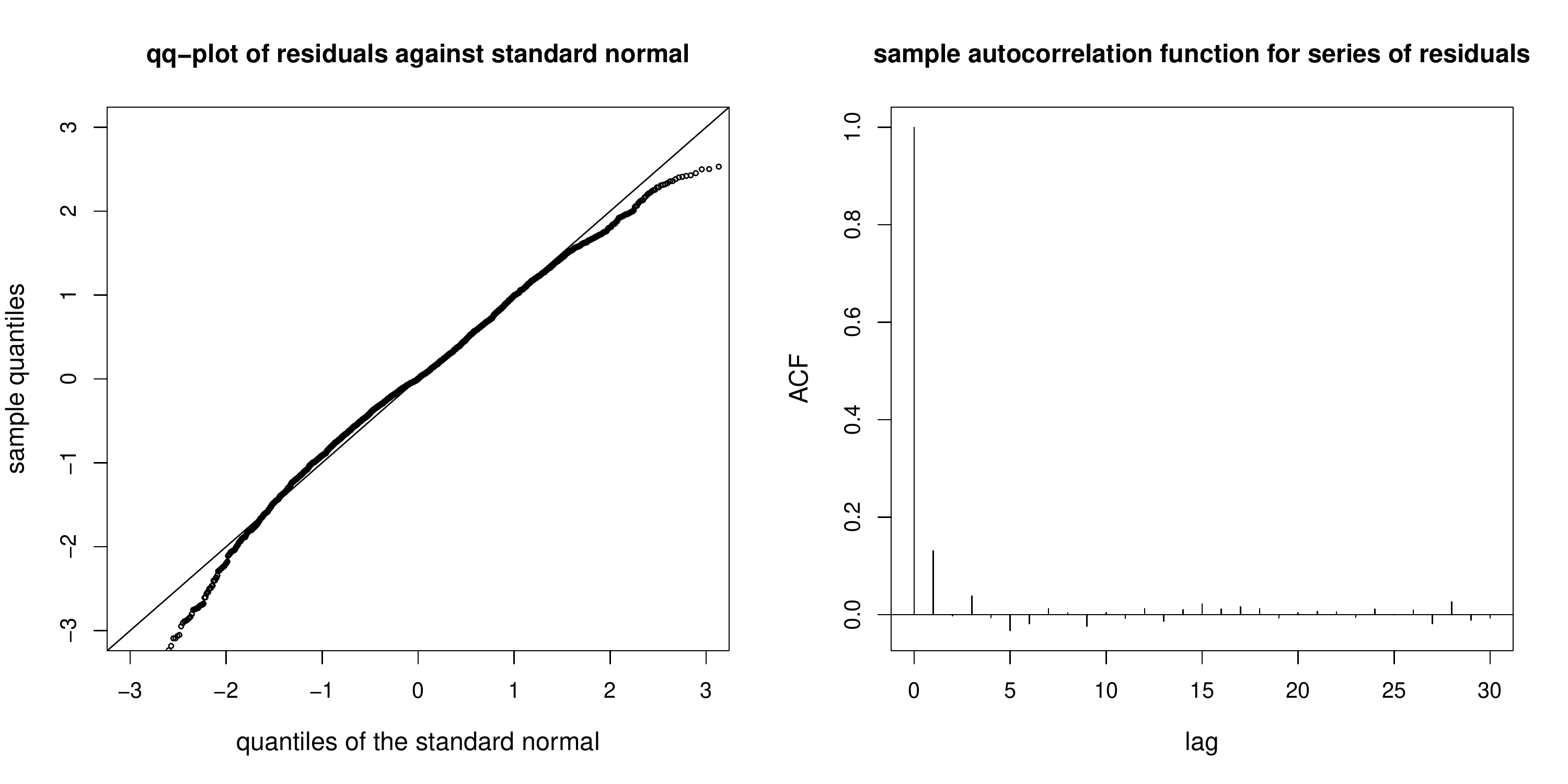}
\end{center}
\vspace{-1em}
\caption{\it Quantile-quantile plot of the one-step-ahead forecast pseudo-residuals obtained for the fitted parametric 3-state model against the standard normal distribution (left panel), and sample autocorrelation function of the series of residuals (right panel).} \label{qq3}
\end{figure}



% ...



% ...



% ...

\end{document}